\documentclass{PoS}
\usepackage{amsmath}
\usepackage{rsfs}

\title{Gribov-Zwanziger horizon condition, ghost and gluon propagators and Kugo-Ojima confinement criterion}

\ShortTitle{Gribov-Zwanziger horizon condition}

\author{\speaker{Kei-Ichi Kondo}\thanks{On sabbatical leave of absence from Department of Physics, Graduate School of Science, 
Chiba University, Chiba 263-8522, Japan.
}\\
Department of Physics, University of Tokyo, 
Tokyo 113-0033, Japan\\
Department of Physics, Chiba University,  
Chiba 263-8522, Japan\\
        E-mail: \email{kondok@faculty.chiba-u.jp}}

\abstract{
We reexamine the conventional arguments concerned with the scaling and decoupling solutions for the ghost and gluon propagators in the Landau gauge Yang-Mills theory.
We point out a few issues to be clarified, which seems to be overlooked in the previous investigations in this field, in the fully non-perturbative treatment.  
We propose a trick which enables one to incorporate the Gribov horizon directly into the self-consistent Schwinger-Dyson equation in the gauge-fixed Yang-Mills theory, using the Gribov-Zwanziger framework with the horizon term. We obtain the following results, irrespective of the choice of the horizon term.
(i) We find that there exists a family of solutions parameterized by one-parameter $w_R(0)$ which was assumed to be zero implicitly. 
The family includes both the scaling and decoupling solutions, and specification of the parameter  discriminates between them. 
(ii) We observe that the inclusion of the horizon term cancels the ultraviolet divergence in the Schwinger-Dyson equation for the ghost propagator and  the resulting  (non-perturbative) self-consistent solution becomes ultraviolet finite.  In other words, the horizon condition interpolates between the infrared behavior and the ultraviolet one.  This leads to the incompatibility of the  conventional multiplicative renormalization scheme  with the Gribov horizon condition. 
}

\FullConference{International Workshop on QCD Green Functions, Confinement, and Phenomenology - QCD-TNT09\\
		 September 07 - 11 2009\\
		 ECT Trento, Italy}

\begin{document}

\section{Introduction}

We consider the SU(N) Yang-Mills theory in $D$-dimensional Euclidean space. 
A naive definition of a quantum Yang-Mills theory in the Landau gauge
$\partial \mathscr{A}=0$ is given by the functional integral:
\begin{equation}
 Z_{\rm YM}  := \int [d\mathscr{A}] \delta(\partial \mathscr{A}) \det (-\partial D[\mathscr{A}]))\exp \{ -S_{YM}[\mathscr{A}]  \} 
 , 
\end{equation}
where $S_{YM}$ is the Yang-Mills action and $-\partial D[\mathscr{A}]$ is the Faddeev-Popov operator.  
For this formula to be correct, each gauge orbit must intersect the gauge fixing hypersurface $\Gamma := \{ \mathscr{A}; \partial \mathscr{A}=0 \}$ once and a unique representative from each gauge orbit can be chosen by imposing  $\partial \mathscr{A}=0$.
However, it has been shown that there exist many representatives obtained as intersections of a gauge orbit with the gauge fixing hypersurface $\Gamma$, called the Gribov copies \cite{Gribov78}.

In order to avoid the Gribov copies, Gribov \cite{Gribov78} proposed to restrict the functional integral to the 1st Gribov region $\Omega$: 
\begin{equation}
 Z_{\rm Gribov}  := \int_{\Omega} [d\mathscr{A}] \delta(\partial \mathscr{A}) \det (-\partial D[\mathscr{A}]))\exp \{ -S_{YM}[\mathscr{A}]  \} 
  ,
\end{equation}
where the Gribov region defined by
\begin{equation}
 \Omega := \{ \mathscr{A}  ; \partial \mathscr{A} = 0 ,  \ -\partial D[\mathscr{A}] > 0 \} \subset \Gamma  
\end{equation}
  is a bounded and convex region including the origin $\{ \mathscr{A}=0\}$. In fact,  $-\partial_\mu D_\mu[\mathscr{A}=0] = -\partial_\mu  \partial_\mu > 0$, i.e., $\{ \mathscr{A}=0\} \in \Omega$.
The boundary of $\Omega$ is called the Gribov horizon:
\begin{equation}
 \partial \Omega := \{ \mathscr{A} ; \partial \mathscr{A} = 0 , \ -\partial D[\mathscr{A}] = 0 \} 
  .
\end{equation}
Then he predicted that the resulting Green functions exhibit unexpected behavior in the deep infrared (IR) region, which plays the essential role in confinement.
The gluon 2-point function (full or complete propagator) 
\begin{equation}
 D_{\mu\nu}^{AB}(k) := \delta^{AB} \left[ \left(\delta_{\mu\nu}-\frac{k_\mu k_\nu}{k^2}  \right) \frac{F(k^2)}{k^2} + \frac{\alpha}{k^2} \frac{k_\mu k_\nu}{k^2} 
\right] \quad (\alpha=0) ,
\end{equation}
and the ghost propagator  
\begin{equation}
 G^{AB}(k) := - \delta^{AB} \frac{G(k^2)}{k^2}
 ,
\end{equation}
with the free case:  
$
F(k^2) \equiv 1
$
and
$
G(k^2) \equiv 1
$ 
behave in the deep infrared (IR) region  $k^2 \ll 1$ as
\begin{equation}
  \frac{F(k^2)}{k^2} \sim  \frac{k^2}{(k^2)^2+M^4} \downarrow 0 , \quad  G(k^2)  \sim \frac{M^2}{k^2} \uparrow \infty \quad (k^2  \downarrow 0) 
 , 
\end{equation}
where  $M$ is a constant with mass dimension called the Gribov mass. 
This power like behavior should be compared with the ultraviolet (UV)  behavior with the logarithmic corrections given by one-loop resumed perturbation or one-loop renormalization group:
\begin{equation}
  F(k^2) \sim \left( \ln \frac{k^2}{\Lambda^2} \right)^\gamma , \quad G(k^2) \sim \left( \ln \frac{k^2}{\Lambda^2} \right)^\delta,\quad  \gamma = - \frac{13}{22}, \quad \delta = - \frac{9}{44}
 . 
\end{equation}

The Gribov prediction was investigated more elaborately  by solving the coupled Schwinger-Dyson (SD) equation for the gluon and ghost propagators where  
the dressing functions $F$ and $G$ are assumed to obey the power behavior with exponents $\alpha$ and $\beta$ respectively: 
\begin{align}
  F(k^2) = A \times (k^2)^{\alpha}, \quad G(k^2) = B \times (k^2)^{\beta} , \quad
  \alpha+2\beta=0, \quad 0 < A, B< \infty 
   .
\end{align}
It has been shown that the scaling relation $\alpha+2\beta=0$ holds and hence a single exponent $\kappa$ is enough for characterizing the IR behavior: 
\begin{equation}
\alpha = 2\kappa >1, \quad \beta= -\kappa <0, \quad
  \  1/2<\kappa<1 \quad (\text{Gribov} \ \kappa=1 ).
\end{equation}
The Gribov prediction corresponds to the limiting value $\kappa=1$.
This result, gluon suppression and ghost enhancement in the IR region, leads to the running (ghost-antighost-gluon) coupling constant with a non-trivial IR fixed point:
\begin{align}
    g^2(k) :=  g^2 F(k^2)G^2(k^2)
  & \rightarrow (0 <) g^2 AB^2 (<\infty) 
 \quad (k^2 \rightarrow 0)  
  .
\end{align}
This solution is called the \textit{scaling solution}.
\textit{The gluon propagator $F(k^2)/k^2$ vanishes in the IR limit $k^2  \downarrow 0$, while the ghost propagator becomes more singular than the free case in the IR region, or the ghost dressing function $G(k^2)$ diverges}, i.e., 
\begin{equation}
 G(0)=\infty .
\end{equation} 
See the excellent review by Alkofer and von Smekal \cite{AS01} for details.

This IR behavior was considered to be reasonable from the viewpoint of color confinement.
Due to Kugo and Ojima \cite{KO79}, all color non-singlet objects can not be observed or confined, in other words, only color singlet objects are observed, if a criterion  $u(0)=-1$ is satisfied in the Lorentz covariant gauge 
(a sufficient condition for color confinement). 
\footnote{
Note that the Kugo-Ojima theory for color confinement   does not take into account the Gribov problem and is based on the usual BRST formulation where the exact color symmetry and the well-defined BRST charge are assumed.  
}
It is claimed in \cite{Kugo95} that \textit{in the Landau gauge, the Kugo-Ojima criterion for color confinement $u(0)=-1$ is equivalent to the  divergent ghost dressing function $G(0)=\infty$ }, since  in the Landau gauge $G(0)$ is related to $u(0)$ according to 
\begin{equation}
G(0)=[1+u(0)]^{-1}
 .
 \label{Kugo}
\end{equation}
Until 2006, it seemed that the scaling solution has been confirmed by the SD equation, the functional renormalization group equation and numerical simulations on lattice.  This lead to the \textit{ghost dominance  picture for color confinement}.  This strategy is called the Gribov-Zwanziger/Kugo-Ojima color confinement scenario.

So far so good. 
However,  these results are questioned by the Orsay group at Universite de Paris Sud.
By careful analyses of the SD equation,  so-called \textit{the decoupling solution} was discovered \cite{Boucaudetal08}: 
\begin{align}
 & F(k^2) = A^\prime \times (k^2)^{ \alpha^\prime }, \quad G(k^2) = B^\prime \times (k^2)^{ \beta^\prime } , \quad
   \alpha^\prime=1, \quad  \beta^\prime=0 , \quad 0 < A^\prime, B^\prime< \infty  ,
\end{align}
which leads to the running coupling going to zero in the IR limit:
\begin{equation}
    g^2(k) :=  g^2 F(k^2)G^2(k^2)
    \sim  g^2 A^\prime B^\prime{}^2 k^2  \rightarrow 0
 \quad (k^2 \rightarrow 0) , 
\end{equation}
although its possibility was mentioned also in \cite{LS02}.
Moreover, reexaminations of numerical simulations on large lattices \cite{SIMPS06,BMMP08,BIMPS09,CM07,CM08,OS08,SS08}, functional renormalization group equation \cite{FMP08} and other methods \cite{ABP08,Dudaletal08} seem to support this result.
\textit{The decoupling solution implies that the gluon propagator goes to the non-zero and finite constant in the IR limit, while  the ghost propagator behaves like free, namely, the ghost dressing function G(0) is non-zero and finite in the IR limit}:
\begin{equation}
 0<G(0)<\infty . 
\end{equation}
In the decoupling solution, the gluon decouples below its mass scale and the ghost is still dominant, although the ghost dominance in the decoupling solution is weaker than that in the scaling solution.

\section{Exact relationship between the ghost dressing function and the Kugo-Ojima parameter}

\begin{figure}[hptb]
\begin{center}
\includegraphics[width=6.0in]{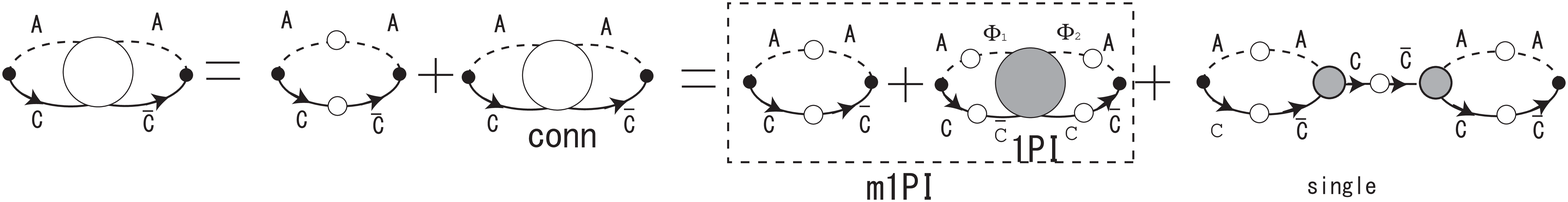}
\end{center}
\caption{Diagrammatic representation of 
$\langle   (g \mathscr{A}_\mu \times \mathscr{C})^A (g \mathscr{A}_\nu \times  \bar{\mathscr{C}})^B \rangle_k$,
$\langle   (g \mathscr{A}_\mu \times \mathscr{C})^A (g \mathscr{A}_\nu \times  \bar{\mathscr{C}})^B \rangle_k^{\rm conn}$, 
$\langle   (g \mathscr{A}_\mu \times \mathscr{C})^A (g \mathscr{A}_\nu \times  \bar{\mathscr{C}})^B \rangle_k^{\rm 1PI}$
and 
$\langle   (g \mathscr{A}_\mu \times \mathscr{C})^A (g \mathscr{A}_\nu \times  \bar{\mathscr{C}})^B \rangle_k^{\rm m1PI}$.
}
\label{fig:4p-def-diagram}
\end{figure}

We wish to point out that the relation (\ref{Kugo}) used in the above discussion  is not precise, since the exact relationship between the ghost dressing function and the Kugo-Ojima parameter is \cite{Kondo09a,Kondo09c}
\begin{equation}
G(0)=[1+u(0)+w(0)]^{-1}
 .
 \label{G}
\end{equation}
Therefore, \textbf{the Kugo-Ojima criterion $u(0)=-1$ is equivalent to the scaling solution $G(0)=\infty$, if and only if $w(0)=0$}. 
In fact, the relationship (\ref{G}) is a special case of the relationship between the ghost dressing function $G(k^2)$ and the Kugo-Ojima function $u(k^2)$ with an additional function $w(k^2)$:
\begin{equation}
  \quad 
  G^{-1}(k^2) =  {1} + u(k^2)+w(k^2) 
  .
\end{equation}
This identity was first derived in \cite{Kugo95} and confirmed in \cite{Kondo09a,Kondo09c}, while it was shown also in \cite{GHQ04} based on a different method.
Here two functions $u$ and $w$ are defined from the modified 1-particle irreducible (m1PI) part as
\begin{equation}
 \lambda_{\mu\nu}^{AB}(k) := \langle   (g \mathscr{A}_\mu \times \mathscr{C})^A (g \mathscr{A}_\nu \times  \bar{\mathscr{C}})^B \rangle_k^{m1PI}
=  \left[ \delta_{\mu\nu} u(k^2) + \frac{k_\mu k_\nu}{k^2} w(k^2) \right] \delta^{AB} 
 ,
\end{equation}
where $u(k^2)$ agrees with the Kugo-Ojima function usually defined by
\footnote{This definition could be reconsidered from the viewpoint of renormalizability.} 
\begin{equation}
  \langle  (D_\mu \mathscr{C})^{A}  (g\mathscr{A}_\nu \times \bar{\mathscr{C}})^{B}  \rangle_{k} 
:= \left(  \delta_{\mu\nu} - \frac{k_\mu k_\nu}{k^2} \right) \delta^{AB} u(k^2)   
 . 
\end{equation}
The m1PI part is defined from the two-point function of the composite operators (See Fig.~\ref{fig:4p-def-diagram})
\begin{equation}
   \langle   (g \mathscr{A}_\mu \times \mathscr{C})^A (g \mathscr{A}_\nu \times  \bar{\mathscr{C}})^B \rangle_k
=    \lambda_{\mu\nu}^{AB}(k)
+  \Delta_{\mu\nu}^{AB}(k)
 ,
\end{equation}
where
\begin{align}
\lambda_{\mu\nu}^{AB}(k) :=&  \langle   (g \mathscr{A}_\mu \times \mathscr{C})^A (g \mathscr{A}_\nu \times  \bar{\mathscr{C}})^B \rangle_k^{\rm m1PI}
,
\nonumber\\
 \Delta_{\mu\nu}^{AB}(k)   :=&  \langle (g \mathscr{A}_\mu \times \mathscr{C})^A \bar{   \mathscr{C}}^C \rangle_k^{\rm 1PI} 
 \langle \mathscr{C}^C \bar{\mathscr{C}}^D \rangle_k 
 \langle \mathscr{C}^D (g \mathscr{A}_\nu \times \bar{   \mathscr{C}})^B \rangle_k^{\rm 1PI}   
   .
\end{align}



We should consider the renormalized version which will be discussed later:
\begin{equation}
  G_R^{-1}(k^2) =  {1} + u_R(k^2)+w_R(k^2) 
  .
  \label{Guw}
\end{equation}
Here we understand that the renormalized quantities $G_R$, $u_R$, and $w_R$ are free from the ultraviolet divergence. 
Note that $w_R(0) \ne 0$ means the existence of the massless pole in the correlation function $\lambda_{\mu\nu}(k)$. 
Indeed, $w_R(0) = 0$ is true at the tree level and holds in perturbation theory, since the massless pole coming from the elementary Faddeev-Popov ghost was already removed in the definition of the 1PI function $\lambda_{\mu\nu}(k)$, see Fig.~\ref{fig:4p-def-diagram}. 
Therefore, within the perturbation theory (up to a finite order), the scaling solution can be compatible with the Kugo-Ojima criterion:
\begin{equation}
  G_R(0) = \infty, \quad u_R(0)=-1 , \quad (w_R(0) = 0).
\end{equation}
However, the Kugo-Ojima criterion $u_R(0)=-1$ can not be realized in perturbation theory, since it is a quantity of order $g^2$. 
Therefore, we must reexamine the whole strategy of GZ/KO from a fully non-perturbative point of view. 
For $w_R(0) \ne 0$ to be realized, a massless boundstate must be formed anew in the channel by the non-perturbative effect and the massless pole must be generated. 
However, the existence of such a pole would not mean the appearance of the physical massless particle in the spectrum, since it is not a gauge-invariant object. 
Therefore, $w_R(0) \ne 0$ does not contradict with the experiment. 
This consideration leads to a possibility of the decoupling solution satisfying the Kugo-Ojima criterion for color confinement. 
\begin{equation}
 0 < G_R(0) < \infty, \quad u_R(0)=-1 , \quad (w_R(0) \ne 0).
\end{equation}
A new possibility of $w_R(0) \ne 0$ should be examined from a non-perturbative point of view, as pointed out in \cite{Kondo09d}.  
We can consider other possible cases which must be compatible with (\ref{Guw}).

\section{The horizon condition and horizon terms}

In order to see the effect of the existence of the Gribov horizon,  
we make use of the Gribov-Zwanziger theory \cite{Zwanziger89,Zwanziger92,Zwanziger93}:   
\begin{equation}
 Z_{\rm GZ} := \int [d\mathscr{A}] \delta (\partial^\mu \mathscr{A}_\mu) \det (K[\mathscr{A}]) \exp \left\{ -S_{YM}[\mathscr{A}] - \gamma \int d^D x h(x) \right\} 
 , 
 \label{YM1}
\end{equation}
where $K$ is the Faddeev-Popov operator $K[\mathscr{A}]:=-\partial_\mu D_\mu[\mathscr{A}]=-\partial_\mu (\partial_\mu+g \mathscr{A}_\mu \times)$ and $h(x)=h[\mathscr{A}](x)$ is the Zwanziger \textit{horizon function}.
Here the parameter $\gamma$ called the \textit{Gribov parameter} is determined by solving a gap equation, commonly called the \textit{horizon condition}: for $D$-dimensional Euclidean SU(N) Yang-Mills theory,
\begin{equation}
 \langle h(x) \rangle_{\rm GZ} = (N^2-1)D .
\end{equation}
The horizon function plays the role of restricting the integration region inside the Gribov horizon $\partial \Omega$, since the horizon function has the property:
\begin{equation}
 \mathscr{A} 
\rightarrow \partial \Omega 
\Longrightarrow  K[\mathscr{A}] \downarrow 0 
\Longrightarrow  \int d^D x h[\mathscr{A}](x) \uparrow \infty 
\Longrightarrow \exp \{  - \gamma \int d^D x h(x) \} \downarrow 0 \ (\gamma >0).
\end{equation}
However, the explicit form of the horizon function respecting exactly the original definition  is not known and the choice of the horizon function is not unique at present. 
The first proposal of the horizon function is \cite{Zwanziger89}  
\begin{equation}
 h(x) 
=  \int d^Dy gf^{ABC} \mathscr{A}_\mu^{B}(x) (K^{-1})^{CE}(x,y) gf^{AFE} \mathscr{A}_\mu^{F}(y)
 . 
 \label{h1}
\end{equation}
The second proposal is \cite{Zwanziger93}
\begin{equation}
  h(x) =  \int d^Dy D_\mu[\mathscr{A}]^{AC}(x) (K^{-1})^{CE}(x,y) D_\mu[\mathscr{A}]^{AE}(y) .
  \label{h2}
\end{equation}
In any case, inclusion of the  horizon term makes the theory non-local.

It should be remarked that the horizon term does not uniquely fix the gauge, since there are still Gribov copies inside the 1st Gribov region. 
Recently, a one-parameter family of correlation functions are constructed in lattice gauge theory  distinguished by a second gauge parameter $B$ (Landau-$B$ gauge) \cite{Maas09}. This uniquely specifies a representative from a gauge orbit and no further freedom in choosing a Gribov copy.

\section{Schwinger-Dyson equation with the horizon condition}

The SD equations for Green functions do not change their form even in the presence of the Gribov horizon, since the integrand of the functional integration formula for Green functions vanishes at the Gribov horizon as the boundary of the functional integration region due to the Faddeev-Popov operator. 
Hence, the solutions of the SD equation include both the solution with the Gribov horizon and the solution without restriction.   
Remarkably, it has been shown  \cite{FMP08,FMP07} that the set of solutions of  the coupled SD equation for gluon and ghost propagators is uniquely determined once a boundary value $G(0)$ is given, corresponding to the scaling solution for $G(0)=\infty$ and the decoupling solution for $0<G(0)<\infty$. 
However, it is not yet examined how these solutions are related to the Gribov horizon.

The Schwinger-Dyson (SD) equation for the ghost propagator $\langle   \mathscr{C}^A \bar{\mathscr{C}}^B \rangle_k$ in   momentum space is written in the following form. 
We follow the notation of \cite{Kondo09a,Kondo09c}.
\begin{equation}
   \langle   \mathscr{C}^A \bar{\mathscr{C}}^B \rangle_k^{-1} = - \delta^{AB} k^2 
-i \frac{k^\mu}{k^2} \langle (g \mathscr{A}_\mu \times \mathscr{C})^A \bar{   \mathscr{C}}^B \rangle_k^{\rm 1PI} 
 ,
\end{equation}
which  is obtained as the Fourier transform of  
\begin{equation}
  0 =  - \langle  (\partial_\mu D_\mu[\mathscr{A}] \mathscr{C})^A(x) \bar{\mathscr{C}}^B(y) \rangle   + \delta^{AB}  \delta^D(x-y) 
 .
\end{equation} 
In the Gribov theory, this is derived from the identity:
\begin{equation}
 0 = \int_{\Omega} [d\mathscr{A}] [d\mathscr{B}][d\mathscr{C}][d\bar {\mathscr{C}}] 
\frac{\delta}{\delta \bar{\mathscr{C}}^A(x)} \left[  e^{ -S_{\rm YM}^{\rm tot}  } \bar{\mathscr{C}}^B(y) \right]
 , 
\end{equation}
while in the Gribov-Zwanziger theory, the same form is obtained   from 
\begin{equation}
 0 = \int [d\mathscr{A}] [d\mathscr{B}][d\mathscr{C}][d\bar {\mathscr{C}}] 
\frac{\delta}{\delta \bar{\mathscr{C}}^A(x)} \left[  e^{ -S_{\rm YM}^{\rm tot} - \gamma \int d^D x h(x) } \bar{\mathscr{C}}^B(y) \right]
 , 
\end{equation}
where
\begin{align}
 S_{\rm YM}^{\rm tot}  :=& S_{\rm YM} + S_{\rm GF+FP} ,
 \nonumber\\
 S_{\rm YM} :=&  \int d^Dx \frac14 \mathscr{F}_{\mu\nu}  \cdot \mathscr{F}_{\mu\nu} ,
\quad 
 S_{\rm GF+FP} :=  \int d^Dx  \left\{ \mathscr{B} \cdot \partial_\mu \mathscr{A}_\mu 
+i \bar {\mathscr{C}} \cdot \partial_\mu D_\mu \mathscr{C} \right\} 
 ,
\end{align}
and the dot and the cross are defined as 
$
 \mathscr{A} \cdot \mathscr{B} := \mathscr{A}^A \mathscr{B}^A
$
and
$
 (\mathscr{A} \times \mathscr{B})^A := f^{ABC} \mathscr{A}^B \mathscr{B}^C 
$.

In \cite{Kondo09d}, we have given a trick to incorporate the horizon condition into the SD equation of the ghost propagator, which enables us to distinguish the solution associated with the particular choice of the horizon term and to discriminate the scaling and decoupling solutions. 
Following the idea of Gribov \cite{Gribov78}, we incorporate the horizon condition 
$
 \langle  h(0) \rangle = (N^2-1)D
$
into the SD equation. 
By substituting the horizon condition into the free (tree) part of the SD equation, the SD equation for the ghost dressing function reads
\begin{equation}
  G^{-1}(k^2) =   \frac{\langle h(0) \rangle}{(N^2-1)D}  + u(k^2)+w(k^2) 
  ,
  \label{SDwHa}
\end{equation}
which is equivalent to the SD equation for the ghost propagator:
\begin{equation}
   \langle   \mathscr{C}^A \bar{\mathscr{C}}^B \rangle_k^{-1} = - \delta^{AB} k^2 
 \frac{\langle h(0) \rangle}{(N^2-1)D} 
-i \frac{k^\mu}{k^2} \langle (g \mathscr{A}_\mu \times \mathscr{C})^A \bar{   \mathscr{C}}^B \rangle_k^{\rm 1PI} 
 .
  \label{SDwHb}
\end{equation} 

Advantages of the SD equation with the insertion of the horizon condition is as follows.
\begin{enumerate}
\item
In order to obtain the scaling solution, the constant terms must cancel exactly or disappear at   $k=0$ on the right-hand side of the SD equation. 
This is what implicitly assumed, but not stated explicitly, 
as pointed out by \cite{Boucaudetal08}.
Otherwise,  the decoupling solution is obtained. 
The SD equation (\ref{SDwHa}) with a horizon condition can be used to discriminate between the scaling and the decoupling solutions as a result of the existence of the Gribov horizon. 

\item
The renormalized version of the SD equation still contains the UV divergence coming from the integration over the internal momenta in the self-energy part even after Green functions (the ghost propagator, gluon propagator and the vertex functions) are replaced by the renormalized ones.  
If we include the horizon term, this UV divergence cancels between the self-energy term and the horizon term and the SD equation becomes a self-consistent equation to yield finite Green functions. 
\end{enumerate}

 We show that both horizon terms allow the existence of one-parameter family of solutions parameterized by a real number $w_R(0)$ which has been assumed implicitly to be zero $w_R(0)=0$ in the previous investigations.  Therefore, $w_R(0)$ plays the role of an additional non-perturbative gauge parameter which uniquely specifies the solution. 
 We consider both the unrenormalized and renormalized versions of the proposed SD equation with the horizon condition being included.

\section{Ultraviolet finiteness of the self-consistent solution of the Schwinger-Dyson equation with the horizon condition}

\subsection{The first horizon term}

The first horizon term (\ref{h1}) yields \cite{Kondo09a,Kondo09c}  
\begin{align}
 \langle  h(0) \rangle
  =& - \lim_{k \to 0}  \langle   (g \mathscr{A}_\mu \times \mathscr{C})^A (g \mathscr{A}_\mu \times  \bar{\mathscr{C}})^A \rangle_k
  \nonumber\\
=& 
 - (N^2-1)  \left\{ Du(0) +w(0)- G(0) [u(0)+w(0)]^2  \right\}  
 ,
 \label{hc1}
\end{align} 
and
\begin{equation}
  \frac{\langle  h(0) \rangle}{(N^2-1)D}
= -u(0)-\frac{w(0)}{D}+ \frac{G^{-1}(0)-2+G(0)}{D}
 .
\end{equation}
Then the SD equation reads 
\begin{equation}
  G^{-1}(k^2) =     \frac{G^{-1}(0)-2+G(0)}{D} -\frac{w(0)}{D} -u(0) + u(k^2)+w(k^2) 
  .
\end{equation}
Then, it is observed that the  term $-u(0)$ coming from the horizon condition exactly cancels the self-energy term $u(k^2)$ at $k=0$ in the right-hand side of the SD equation.  
In the deep IR limit $k=0$,  therefore, we have
\begin{equation}
  G^{-1}(0) =  
 \frac{G^{-1}(0)-2+G(0)}{D}
+ \left(  - \frac{1}{D}  + 1  \right) w(0) 
  .
  \label{ghostSDEda2}
\end{equation}
By solving this equation for $G(0)$:
$
  G^{2}(0) -[2-(D-1)w(0)]G(0) + 1-D = 0 ,
$
we have
\begin{equation}
G(0) = 1 -(D-1)w(0)/2+ \sqrt{[1-(D-1)w(0)/2]^2-1+D} > 0
 ,
\end{equation} 
and $u(0) =  -1-w(0)+G^{-1}(0)$ reads 
\begin{equation}
u(0) 
= -1-w(0)- \frac16 \left\{ 2-3w(0)-\sqrt{12+[2-3w(0)]^2} \right\}
 .
\end{equation} 
This implies  that \textit{the horizon condition determines the boundary value $G(0)$ in the ghost SD equation. 
Consequently, we have one-parameter family of solutions parameterized by $w(0)$.}



We consider $G(0)$ and $u(0)$ as  functions of $w(0)$.
Both $G(0)$ and $u(0)$ are monotonically decreasing functions in $w(0)$;    $G(0), u(0) \rightarrow +\infty$ as $w(0) \rightarrow -\infty$, while $G(0) \rightarrow 0$ and $u(0) \rightarrow -5/3$ as $w(0)  \rightarrow +\infty$.  
Thus the scaling solution $G(0)=+\infty$ is obtained only when $w(0)=-\infty$. 
Otherwise $w(0)>-\infty$, the decoupling solution $0< G(0) < \infty$ is obtained.

Using a special value as an additional input $w(0)=0$ which is assumed in \cite{Kugo95,Kondo09a} or suggested by an independent analysis \cite{ABP09}, 
 $G(0)$ is determined self consistently by solving the above SD equation as \cite{Kondo09a,Kondo09c}
\begin{equation}
G(0) = 1 + \sqrt{D} > 0,
\quad
u(0)=(-D \pm \sqrt{D})/(D-1)
 .
\end{equation} 
In particular, for $D=4$, 
\begin{equation}
G(0) = 3 > 0,
\quad
u(0)=-2/3 \quad (D=4)
 .
\end{equation} 
This agrees with the current data of numerical simulations \cite{FN07,Sternbeck06}.

We proceed to the treatment of the ultraviolet divergence. The naive SD equation without the horizon condition reads
\begin{equation}
  G^{-1}_{\Lambda}(k^2) =    1 + u_{\Lambda}(k^2)+w_{\Lambda}(k^2) 
  .
\end{equation}
Here, in order to avoid the ultraviolet divergence in the self-energy part $u(k^2)+w(k^2)$, the UV cutoff $\Lambda$ has been introduced thereby to make the self-energy part $u_{\Lambda}(k^2)+w_{\Lambda}(k^2)$ finite. 
Consequently,  $G$ must depend on $\Lambda$, i.e.,  $G_{\Lambda}(k^2):=G(k^2,\Lambda)$. 
In this sense, the above SD equation is an \textit{unrenormalized} version.

It is pointed out in \cite{Boucaudetal09}   that by writing $u(0)$ and $w(0)$ as function of $G(0)$, one finds $u_{\Lambda}(0) \rightarrow +\infty$ and $w_{\Lambda}(0) \rightarrow -\infty$  such that $u_{\Lambda}(0)+w_{\Lambda}(0) \rightarrow -1$, provided that  $G_{\Lambda}(0) \rightarrow \infty$ as $\Lambda \rightarrow \infty$, see Appendix~C of \cite{Kondo09d}.
This analysis uses the first horizon condition and the relation $G(0)^{-1}=1+u(0)+w(0)$. However, the statement $G_{\Lambda}(0) \rightarrow \infty$ as $\Lambda \rightarrow \infty$ is a result of perturbation theory.  This analysis does not use the full information coming from the relation $G(k^2)^{-1}=1+u(k^2)+w(k^2)$  (the Schwinger-Dyson equation for the ghost dressing function)  for the whole momentum region.

In what follows, we show that {\bf $G_{\Lambda}(0)$ remains finite  in a non-perturbative way, even after sending the UV cutoff to infinity $\Lambda \rightarrow \infty$}, once the full information on the whole momentum region is used.  
We consider the SD equation with the   \textit{horizon condition}:
\begin{equation}
  \quad 
  G^{-1}_{\Lambda}(k^2) =    \frac{G^{-1}_{\Lambda}(0)-2+G_{\Lambda}(0)}{D} -\frac{w_{\Lambda}(0)}{D}  
 -u_{\Lambda}(0)  + u_{\Lambda}(k^2)+w_{\Lambda}(k^2) 
  .
  \label{ghostSD-hc}
\end{equation}

If the horizon condition is incorporated into the SD equation, a partial cancellation at $k=0$ occurs between the horizon condition and the ghost self-energy.  This cancellation always occurs for the (multiplicative renormalizable) part coming from $\lambda_{\mu\mu}^{AA}(0)$, while this is not the case for the contribution from the remaining term $\Delta_{\mu\mu}^{AA}(0)$.

The first horizon condition which is non-linear in $G^{-1}$ is not multiplicatively renormalizable \cite{Kondo09d}. Therefore, the ultraviolet divergence can not be removed within the usual renormalization scheme. 
However, a novel situation occurs by introducing the first horizon condition into the SD equation.
By the resulting combination  $u_{\Lambda}(k^2)-u_{\Lambda}(0)$,   the ultraviolet divergence cancels exactly and  the ultraviolet cutoff $\Lambda$ can be sent to infinity to obtain a finite function of $k^2$. 
Then the above SD equation is regarded as a self-consistent equation to give a finite ghost function $G(k^2)=\lim_{\Lambda \rightarrow \infty}G_{\Lambda}(k^2) < \infty$. 
The term $w_{\Lambda}(k^2)$ is finite from the beginning by some reason, although the $\Lambda$ dependence is apparently assumed, see Appendix~B of \cite{Kondo09d}.
In other words, the SD equation is self-organized (in a non-perturbative way) to give a finite result. 
Therefore there is no need for the specific ultraviolet renormalization in the presence of the first horizon term!
Thus, the results obtained in the unrenormalized case hold also after the ultraviolet cutoff $\Lambda$ is send to infinity, as far as the ghost propagator or the dressing function is concerned. 

Even if the ghost dressing function is ultraviolet finite, we have still finite renormalization coming from the choice of the renormalization point, leading to consistent results with numerical simulations. See \cite{Kondo09d}.

\subsection{The second horizon term}

We can perform the similar analysis for the second horizon term. See \cite{Kondo09d} for details.

\section{Conclusion and discussion}

We have discussed how the existence of the Gribov horizon affects the deep infrared behavior of the ghost propagator in the Landau gauge $G=SU(N)$ Yang-Mills theory, using the Gribov-Zwanziger framework with the horizon condition $\langle h(x) \rangle = ({\rm dim}G)D$. 
 Moreover, we have shown how to incorporate the horizon condition into the Schwinger-Dyson equation for the ghost propagator to discriminate between scaling and decoupling. 
 We have examined two horizon conditions derived from two types of horizon terms, both of which were proposed by Zwanziger \cite{Zwanziger89,Zwanziger93}. 
 We have shown that one parameter family of solutions parameterized by $w(0)$ exists in both cases, although some results crucially depend on the choice  of the horizon term adopted. 
 The value $w(0)$ has been assumed implicitly to be zero $w(0)=0$ in previous studies.  
 
Thus the investigation of  $w_R(0)$ is crucial to see which solution is realized.  
An interesting step towards this direction was done in \cite{ABP09}. However, according to our analysis, their result, namely, the  decoupling solution with $w_R(0)=0$ is not compatible with the multiplicative renormalization scheme (without the horizon term): 
\begin{equation}
  G^{-1}(k^2) = Z_{C}^{-1} G^{-1}_R(k^2) 
  \Longleftrightarrow 
 1+ u(k^2)  
= Z_{C}^{-1} [1+ u_R(k^2) ], \quad
 w(k^2) = Z_{C}^{-1}w_R(k^2) ,
\end{equation}
since $w(k^2)$ is ultraviolet finite in contradict to $Z_{C}=\infty$.  This issue should be reexamined and confirmed by further investigations.

We can regard that $w_R(0)$ plays the role of an additional non-perturbative gauge parameter which uniquely specifies the solution from a one-parameter family of solutions including the scaling and decoupling.  
In view of this, the value of $w_R(0)$  itself   might have  no physical meaning. 
In fact, it has been shown \cite{BGP07} that all solutions (decoupling as well as scaling) lead to quark confinement by proving the vanishing of  the Polyakov loop as a gauge-invariant order parameter of quark confinement.
In this sense, discriminating between decoupling and  scaling may not be so important from the physical point of view and main contribution to phenomenological studies comes from the high-momentum region and the intermediate region around 1GeV which is stable irrespective of adopting the scaling or decoupling solution. 
It is shown that both scaling and decoupling solutions do not contradict the general principles of quantum gauge field theories \cite{Kondo04}.

How the existence of the horizon is relevant for color confinement. 
In the Gribov-Zwanziger theory (restricted to the 1st Gribov region), the BRST symmetry is broken by the existence of the horizon. 
$
\mbox{\boldmath $\delta$} S_{\rm GZ}
=\mbox{\boldmath $\delta$} \tilde S_\gamma \ne 0
$.
Nevertheless, there exists  a ``BRST'' like symmetry (without nilpotency \cite{Sorella09}  or with nilpotency \cite{Kondo09b} which leaves the Gribov-Zwanziger action invariant.  
Then we could apply the Kugo-Ojima idea to the Gribov-Zwanziger theory, which opens the path to searching for the modified color confinement criterion {\it a la} Kugo and Ojima.  
On the other hand, defining a non-perturbative BRST transformation without restricting to the first Gribov region will be another interesting possibility to be investigated   \cite{Smekal08}.

\section*{Acknowledgments}
The author would like to thank the organizers: Danielle Binosi, Joannis Papavassiliou, John Cornwall and  Arlene Aguilar for the kind invitation and hospitality at ECT*, Trento. 
This work is financially supported by  Grant-in-Aid for Scientific Research (C) 21540256 from Japan Society for the Promotion of Science
(JSPS).

\end{document}